\documentstyle[epsf,preprint,eqsecnum,aps,rotate]{revtex}
\begin{document}

\global\firstfigfalse
\global\firsttabfalse

%\draft

%\preprint{RU-96-82}

\title{The CP/T Experiment}

\author{Gordon Thomson}
\address{Rutgers University, Piscataway, NJ}

\maketitle

\begin{abstract}

\vspace{-0.5in}

      In this talk I describe a proposed Fermilab Main Injector
experiment to carry out a program of measurements on the physics of
$K^0$ mesons.  The experiment is designed to maximize the interference
between $K_L$ and $K_S$ mesons near their production target, and hence
have excellent sensitivity to CP violation in many decay modes.  The
extremely accurate CP violation measurements we will be able to make
will allow us to test CPT symmetry violation with sensitivity at the
Planck scale.

     The experiment will use an RF-separated $K^+$ beam striking a
target at the entrance to a hyperon magnet to make the $K^0$ beam by
charge exchange.  The decay region, magnetic spectrometer,
electromagnetic calorimeter, and muon detector follow immediately to
observe interference between $K_L$ and $K_S$ near the target.

\end{abstract}

\section{Introduction}

     This talk is a description of an experiment that has been
proposed to run at the Fermilab Main Injector to study CP violation,
test CPT symmetry conservation, and search for rare decays of the
$K_S$ meson.  Here I will only discuss the CPT symmetry
nonconservation search.

     CPT symmetry conservation is a subject under theoretical attack.
Studies of Hawking radiation \cite{kn:hawking} and of string theory
(the leading contender for a unified theory of all four forces of
nature) \cite{kn:stcomment} have shown the CPT theorem
\cite{kn:luders} to be invalid in real life (rather than in the
three-force approximation we call the standard model).  Many
physicists are reluctant to accept the possiblity that CPT symmetry
violation may occur at the Planck scale.  One reason for this
reluctance is that we have, so far, only theoretical hints that this
is the case.  Another reason is the great success of the standard
model.  It may be quite a few years before a theory that unifies all
four of the forces of nature becomes mature enough so that convincing
theoretical statements can be made about the CPT structure of the
world.

     Fortunately we don't have to wait.  The $K_L - K_S$ system
provides a way of testing the validity of CPT symmetry conservation
where it is possible to perform extremely accurate experiments
\cite{kn:schwingenheuer}.  In this document we propose to do an
experiment that will reach the Planck scale.  Finding CPT symmetry
nonconservation would be a major discovery that would change in a
fundamental way how physicists view the world.  If we don't find it we
will strongly constrain several quantum theories of gravity
\cite{kn:ellis} \cite{kn:kostelecky} and provide a powerful benchmark
against which future theories must be measured.

\section{CPT Theory and Phenomenology}

     The CPT theorem \cite{kn:luders} is based on the assumptions of
locality, Lorentz invariance, the spin-statistics theorem, and
asymptotically free wave functions.  All quantum field theories
(including the standard model of the elementary particles) assume CPT
symmetry invariance.

     There is a theoretical hint of the level at which CPT symmetry
might be violated.  This comes from the fact that gravity can't be
consistently included in a quantum field theory, and the proof of the
CPT theorem assumes Minkowski space \cite{kn:wald,kn:luders}.  To include
gravity in a unified theory of all four forces of nature, many
physicists think that a more general theory is needed, which would
have quantum field theory embedded in it.  In this more general theory
the CPT theorem will be invalid.

     One expects to see quantum effects of gravity at what is called
the Planck scale: at energies of $M_{Planck}c^2 = \sqrt{\hbar c^5/G} =
1.2 \times 10^{19}$ GeV, or at distances of the order of $10^{-33}$
cm.  The quantum effects of gravity are expected to be very small in
ordinary processes.  However, in a place where the standard model
predicts a null result, like CPT violation, quantum effects of gravity
would stand out.  Therefore, it would be very interesting to test CPT
symmetry conservation at the Planck scale.

     One might think that string theory, as a candidate for the more
general theory that has quantum field theory embedded in it, would
give us guidance.  CPT conservation is artificially built into string
theories, first by G. Veneziano \cite{kn:banks}.

     Kostelecky and Potting \cite{kn:kostelecky} suggested that
spontaneous CPT violation might occur in string theory; i.e., they put
the CPT violation in the solutions rather than in the equations of
motion.  One of the problems with string theory in general is that
it's not known how to relate string effects at the Planck scale to
effects seen at current accelerator energies, and Kostelecky and
Potting have the same difficulty.  They have tried to remedy this by
writing the most general additions to the Standard Model Lagrangian
that maintain the SU(3) x SU(2) x U(1) effective structure of the
theory but violate CPT symmetry.  This allows them to classify the
various types of CPT violation that might be seen (in the lepton
sector, quark sector, etc.) and have a parameterization that includes
all these effects.  They find that the largest CPT violating effect is
a change in quark propagators that has the opposite sign for
antiquarks.  This leads to a nonzero value of $|M_{K^0} -
M_{\overline{K^0}}|$ coming from indirect CPT violation.  This is much
larger than any direct CPT violation effect.  This is precisely the
signature that this experiment would search for.

     The $K^0 - {\overline K^0}$ system provides us with an incredibly
finely balanced interferometer that magnifies small perturbations such
as CPT violating effects.  It is a natural place to search for CPT
symmetry violation since it exhibits C, P, and CP symmetry violation
(and is the only place to date where CP violation has been seen).  In
the final analysis, the conservation or violation of CPT symmetry is
an experimental question, and the search for this effect is of the
utmost interest.

     In $K^0$ physics, one can observe CPT violating effects through
mixing or decays (called indirect or direct CPT violation
respectively).  In mixing, one introduces a parameter $\Delta$ which
is both CP and CPT violating.  One can also have direct CPT violation.
Eqn. (\ref{eq:kkbar}) shows the mixing of $K_L$ and $K_S$ in terms of
the CP eigenstates $K_1$ and $K_2$.
\begin{eqnarray}
\left\{ \begin{array}{c} K_S = K_1 + (\epsilon + \Delta)K_2 \\
K_L = K_2 + (\epsilon - \Delta)K_1 \end{array}
\right.
\label{eq:kkbar}
\end{eqnarray}

     There are several measurements that would signify CPT violation:
a difference between the phase of $\epsilon$ and the phase of
$\eta_{+-}$, evidence for a non-zero $\Delta$ in the Bell-Steinberger
relation, a difference between the phases of $\eta_{+-}$ and
$\eta_{00}$, or certain interference terms between $K_L$ and $K_S$ in
semileptonic decays.  In this report we will concentrate on the first
two methods, measuring the phase of $\eta_{+-}$ and comparing it to
the calculated value of the phase of $\epsilon$, and evaluating the
Bell-Steinberger relation, since from them we can make the most
accurate measurements.

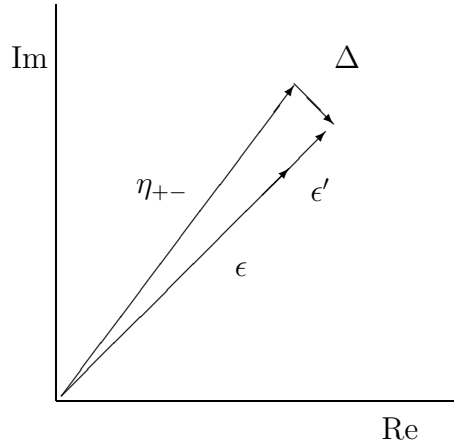
\begin{figure}[bth]
\begin{center}
\begin{picture}(400,200)
\put(120,40){\line(1,0){150}}
\put(250,30){\makebox(0,0){Re}}
\put(120,40){\line(0,1){150}}
\put(110,170){\makebox(0,0){Im}}
\put(122,42){\vector(1,1){100}}
\put(190,90){\makebox(0,0){$\epsilon$}}
\put(198,118){\vector(1,1){10}}
\put(220,120){\makebox(0,0){$\epsilon '$}}
\put(210,160){\vector(1,-1){15}}
\put(230,170){\makebox(0,0){$\Delta$}}
\put(122,42){\vector(3,4){88}}
\put(160,120){\makebox(0,0){$\eta_{+-}$}}
%\put(200,0){\makebox(0,0){Fig. 1.  The Wu-Yang Diagram}}
\end{picture}
\end{center}
\caption{The Wu-Yang Diagram}
\label{fig:wuyang}
\end{figure}

     We now consider the CPT test based on measuring the phase of
$\eta_{+-}$ and calculating the phase of $\epsilon$.  For what follows
we adopt the Wu-Yang phase convention.  Figure \ref{fig:wuyang} shows
the relationships between $\epsilon, \epsilon', \Delta,$ and
$\eta_{+-}$.  $\epsilon '$ and $\Delta$ are shown greatly enlarged for
clarity.

\vspace{0.25in}

The size of $|\epsilon '/\epsilon|$ is of order $10^{-3}$, and the
phase of $\epsilon '$ is very close to that of $\epsilon$, so the
phase of the vector $\epsilon + \epsilon '$ is the same, to good
accuracy, to the phase of $\epsilon$ ($\epsilon'$ is too small to have
an affect on the calculation of the phase of $\epsilon$ at the level
in which we are interested).  We can see from the figure that the
component of $\Delta$ perpendicular to $\epsilon$, $\Delta_{\perp}$,
is
\begin{eqnarray}
\Delta_{\perp} = |\eta_{+-}|(\phi_{+-} - \phi_{\epsilon})
\label{eq:deltaperp}
\end{eqnarray}
where $\phi_{+-}$ ($\phi_{\epsilon}$) is the phase of $\eta_{+-}
(\epsilon)$.  In general, in terms of the elements of the kaon decay
matrix $\Gamma$ and mass matrix $M$, $\Delta$ is given by \cite{kn:leewu}:
\begin{eqnarray}
\Delta = \frac{(\Gamma_{11}-\Gamma_{22}) + i(M_{11}-M_{22})}
              {(\Gamma_S-\Gamma_L) - 2i(M_L-M_S)}
\label{eq:gammam}
\end{eqnarray}

The mass term has a phase perpendicular to $\phi_{SW}$, the superweak
phase, which is defined as $\tan \phi_{SW} =
2(M_L-M_S)/(\Gamma_S-\Gamma_L)$.  $\phi_{SW}$ is approximately equal
to $\phi_{\epsilon}$.  The decay term is parallel to $\phi_{SW}$.  We
can solve Eqns. (\ref{eq:deltaperp}) and (\ref{eq:gammam}) for
$M_{11}-M_{22}$, which is the mass difference between the $K^0$ and
$\overline{K^0}$ mesons, and get an equation which we can use to
search for indirect CPT violation:

\begin{eqnarray}
\frac{|M_{K^0} - M_{\overline{K^0}}|}{M_{K^0}} =
\frac{2(M_L-M_S)}{M_{K^0}} \frac{|\eta_{+-}|}{\sin \phi_{SW}}
|\phi_{+-} - \phi_{\epsilon}|
\label{eq:basiceqn}
\end{eqnarray}

     In Eqn. (\ref{eq:basiceqn}), Nature has been kind: the constant
factors multiplying $|\phi_{+-} - \phi_{\epsilon}|$ are exceedingly
small.  $(M_L-M_S)$ is $10^{-6}$ eV, and when one divides by $M_{K^0}$
the ratio is of order $10^{-15}$. $|\eta_{+-}|$ is of order $10^{-3}$.
The product of all the factors multiplying $|\phi_{+-} -
\phi_{\epsilon}|$ is $4 \times 10^{-17}$.  By the Planck scale we mean
\begin{eqnarray}
\frac{|M_{K^0} - M_{\overline{K^0}}|}{M_{K^0}} =
\frac{M_{K^0}}{M_{Planck}} = 4.1 \times 10^{-20}
\label{eq:planck}
\end{eqnarray}
so a measurement of $|\phi_{+-} - \phi_{\epsilon}|$ accurate to 1
milliradian would test a CPT violating effect at the accuracy of the
Planck scale.

     Some CP/T experiment collaborators were part of Fermilab
experiment E773.  In this experiment we placed the limit (at 90\%
confidence level) \cite{kn:schwingenheuer},
\begin{eqnarray}
\frac{|M_{K^0} - M_{\overline{K^0}}|}{M_{K^0}} < 1.3 \times 10^{-18}
\label{eq:E773}
\end{eqnarray}
so the result of Ref. \cite{kn:schwingenheuer} stands at 31 times the
Planck scale.

     That publication actually compared the phase of $\eta_{+-}$ to
the superweak phase (and stated clearly that in doing so the
assumption was being made that CP violation would not be unexpectedly
large in modes other than $\pi \pi$).  In the calculation of the phase
of $\epsilon$, there are three corrections that should be made to the
superweak phase: from $Im(x)$, the $\Delta S = \Delta Q$ rule
violation parameter, from $Im(\eta_{+-0})$, and from $Im(\eta_{000})$.
Together they have an uncertainty of 2.7 degrees which should be added
in quadrature with the approximately 1 degree accuracy of
Ref. \cite{kn:schwingenheuer}.

     Several CP/T experiment collaborators are part of the KTeV
experiment as well.  There we expect to make an improvement of a
factor of 3 to 5.  In KTeV interference is seen very clearly.  But the
interference term from which $\phi_{+-}$ is measured, $2 |\eta_{+-}|
|\rho| \cos (\Delta m t + \phi_{\rho} - \phi_{+-}) \exp (-t/2
\tau_s)$, is reduced by the regeneration amplitude $|\rho| \simeq
0.03$, and $\phi_{+-}$ and $\phi_{\rho}$ are hard to disentangle.
Using the regeneration method will be difficult beyond the KTeV level
\cite{kn:briere}.

     It should be understood clearly that measuring the phase of
$\eta_{+-}$ and comparing it to the superweak phase does not
constitute a complete test of CPT symmetry conservation: the
corrections to the superweak phase have larger uncertainties than
existing experimental measurements of $\phi_{+-}$.  For example, if a
significant difference between $\phi_{+-}$ and $\phi_{SW}$ were found
in an experiment it would NOT prove that CPT symmetry was violated.
More accurate measurements of $Im(x), Im(\eta_{+-0})$, and
$Im(\eta_{000})$ must be made before this could be proved.  An
interference experiment located just downstream of the production
target is needed for these measurements.  In a regeneration experiment
the interference in $3\pi$ decays is reduced in size by a factor of
$\rho$, the regeneration amplitude, which is about 0.1 at most (at
Main Injector energies), compared to an experiment near the production
target, and it's extremely difficult for a regeneration experiment to
measure $Im(x), Im(\eta_{+-0})$, and $Im(\eta_{000})$ to the required
accuracy.

\section{Two Tests of CPT Symmetry Conservation}

\subsection{The Phase Difference between $\eta_{+-}$ and $\epsilon$}

     After the KTeV experiment we expect to stand an order of
magnitude above the Planck scale.  To close that gap we will want to
do an interference experiment near the kaon production target.  The
interference term is then $2 D |\eta_{+-}| \cos (\Delta m t -
\phi_{+-}) \exp (-t/2 \tau_s)$.  Here $\phi_{+-}$ appears alone, and
$|\rho|$ is replaced with the dilution factor, $D=(K^0-{\overline
K^0})/(K^0+{\overline K^0})$ at the target.  To maximize D and hence
the interference, we choose to make our $K^0$ beam from a $K^+$ beam
by charge exchange.  Then at medium to high Feynman x, $D \simeq 1$.
The charge exchange cross section is large, about 20\% of the total
cross section.  To maximize the flux of $K^+$ made from the 120 GeV/c
protons from the Fermilab Main Injector we choose a $K^+$ momentum of
25 GeV/c.  We would use a hyperon magnet to define the $K^0$ beam,
similar to the one in the Fermilab Proton Center beam line.  In the
calculations described below we assume the use of a vee spectrometer,
a lead glass electromagnetic calorimeter, and a muon detector.

     In Ref. \cite{kn:schwingenheuer} $\phi_{+-}$ was measured to
$1^o$ accuracy.  A CPT-violating mass difference exactly at the Planck
scale would result in $|\phi_{+-} - \phi_{\epsilon}| = 0.06^o$.  We
set ourselves the goal of measuring $\phi_{+-}$ and $\phi_{\epsilon}$
to sufficient accuracy to see such a CPT-violating effect.

     We have calculated the statistical sensitivity of the CPT
measurements assuming that we have a 1 year long run with $3 \times
10^{12}$ protons per pulse at 52\% efficiency.

\begin{figure}[tbh]
\epsfysize=4.0in
\centerline{\epsffile{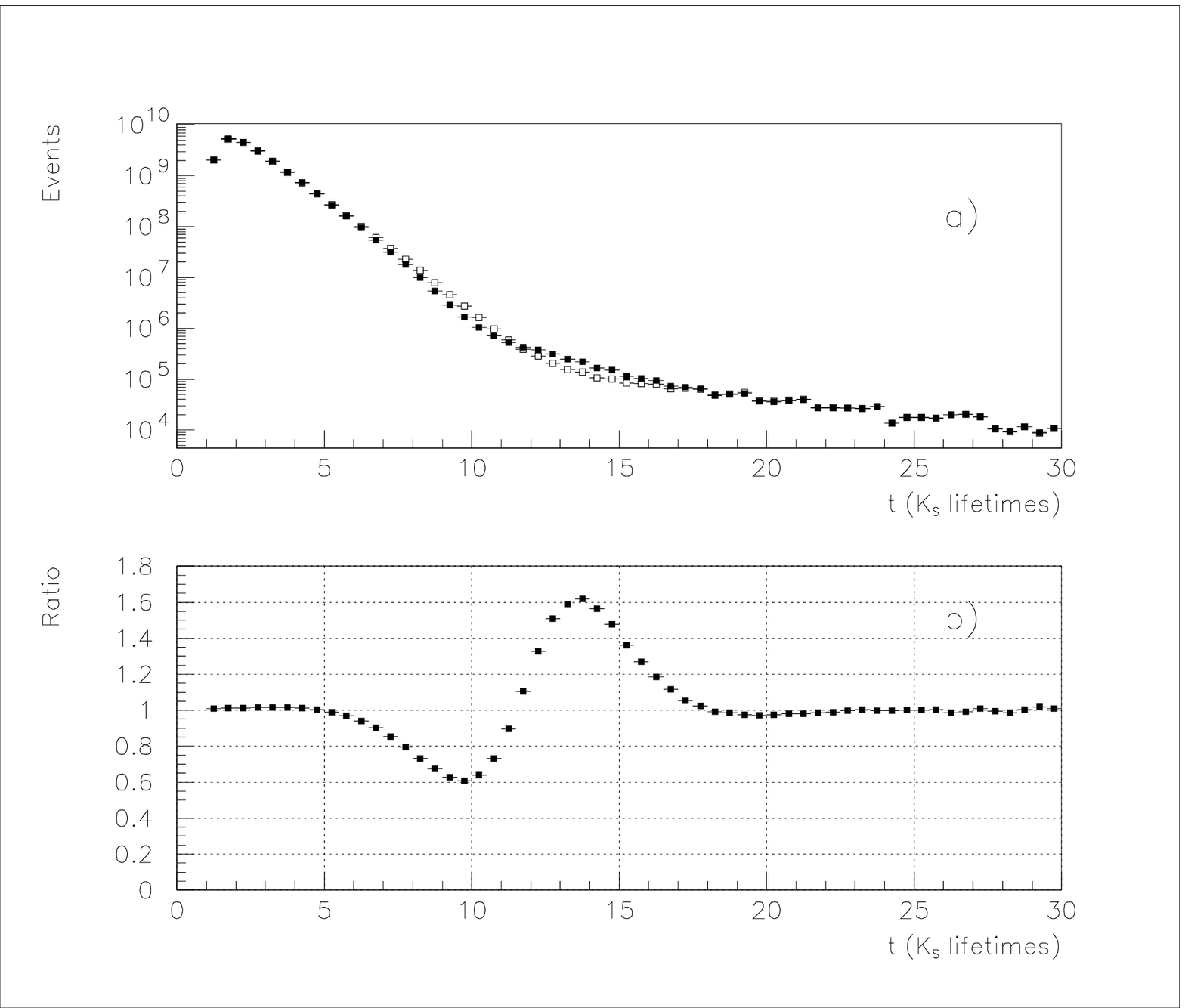}}
\bigskip
\caption{Proper Time Distributions for $\pi^+ \pi^-$ Decays
a) Distributions are shown both with interference (dark squares) and
without (light squares).  b) The ratio of the two distributions in part a).}
\label{fig:tdist}
\end{figure}

     Fig. \ref{fig:tdist} shows the proper time distribution of
accepted events.  The figure shows the actual proper time distribution
and also what the distribution would look like if there were no
interference.  The second part of the figure shows the ratio of those
two curves.  Between 5 and 20 $K_S$ lifetimes the interference is
first a 40\% destructive effect then is a 65\% constructive effect.

     We calculated the distribution of events in momentum and proper
time for the resulting 20 billion events and fit this distribution
using MINUIT, with fitting parameters $|\eta_{+-}|, \phi_{+-}, D$,
three parameters describing the normalization and shape of the kaon
momentum spectrum, $\tau_S$ and $\Delta m$ (the $K_S$ lifetime and the
$K_L - K_S$ mass difference).  The uncertainty that results from this
fit is $0.040$ degrees.  This will meet our goal of testing CPT
symmetry conservation at the Planck scale.  This number $\pm 0.040$
degrees has another meaning: it is the statistical (including fitting)
uncertainty of this measurement, and sets the scale against which all
other aspects of the $|\phi_{+-} - \phi_{\epsilon}|$ measurement
should be compared.

     In this experiment we measure $\phi_{+-}$, but we must also
determine $\phi_{\epsilon}$.  The leading contribution to
$\phi_{\epsilon}$ is the superweak phase, $\phi_{SW}$, given by
$\tan(\phi_{SW}) = 2\Delta m/\Delta \Gamma$.  The superweak phase will
be measured by KTeV to accuracy sufficient for our purposes here.  We
next describe some corrections to this contribution.

      For this experiment, $\epsilon'$ will have no meaningful
effect.  Assuming CPT invariance, the phase of $\epsilon'$ is known
to be $(48 \pm 4)$ degrees \cite{kn:barnett}.  Its magnitude is
unknown, but if we assume it to be the central value from E832 we find
that the maximum possible difference it can provide between
$\phi_{+-}$ and $\phi_{\epsilon}$ is 0.012 degrees, a factor of 5
smaller than the contribution of CPT violation at the Planck scale.

     The full formula for $\phi_{\epsilon}$ is \cite{kn:lavoura}
\begin{eqnarray}
\tan \phi_{\epsilon} = \frac{2 \Delta m}{\Gamma_S - \Gamma_L} \cos
\xi + \frac{\sin \xi}{\delta}
\label{eq:phiep}
\end{eqnarray}
where $\xi = \arg (\Gamma_{12} A_0 {\overline A_0}^*)$ and $\delta
= 2 Re(\epsilon)$.  Here $A_0$ is the isospin 0 part of the $\pi^+
\pi^-$ decay amplitude.  In the Wu-Yang phase convention, $A_0$ is
real, and $\Gamma_{12}$ gives contributions from two sources:
semileptonic decays through $Im(x)$, the $\Delta S = \Delta Q$
violation parameter, and $3\pi$ decays through $Im(\eta_{+-0})$ and
$Im(\eta_{000})$.

     In the standard model we expect $x \simeq 10^{-7}$, which is too
small to affect this experiment, but $Im(x)$ is known experimentally
only to an accuracy of $\pm 0.026$.  This results in an uncertainty in
$\phi_{\epsilon}$ of 1.7 degrees.  To prove that an observed
difference between $\phi_{+-}$ and $\phi_{\epsilon}$ were due to CPT
violation one would have to measure $Im(x)$ about 40 times more
accurately than today's level.  The way we will do this is described
below.

     The contribution to $\phi_{\epsilon}$ from the $3\pi$ modes in the
standard model is 0.017 degrees, which is smaller than the accuracy we
are trying to obtain.  But if one takes into account the current
world's knowledge, the uncertainty these decay modes contribute is 2.2
degrees.  So they also have to be measured better.

     The best experimental approach to measuring these three
quantities, $x, \eta_{+-0}$, and $\eta_{000}$, is the same: choose an
experiment with high dilution factor and observe interference between
$K_L$ and $K_S$ close to the target; i.e. the experiment described
here.  These measurements should be thought of as being an itegral
part of this experiment.  We have performed a calculation of the
sensitivity of this experiment for these quantities, and we estimate
that we can reach at least the required sensitivity.  We conclude that
we can determine $\phi_{\epsilon}$ to the required accuracy.

     We used the same Monte Carlo and fitting programs to estimate the
sensitivity of our experiment to the measurements necessary for the
calculation of $\phi_{\epsilon}$, $Im(x), Im(\eta_{+-0})$, and
$Im(\eta_{000})$, and conclude that we will have the required
sensitivity.  We find that the uncertainty in $Im(x)$ contributes
much more than $Im(\eta_{+-0})$ and $Im(\eta_{000})$ to the
uncertainty in $\phi_{\epsilon}$.

\subsection{CPT Test via the Bell-Steinberger Relation}

     The next test of CPT symmetry conservation comes through an
evaluation of the Bell-Steinberger relation.  Our ability to measure
CP violation parameters (and also $Im(x)$) very accurately will make
it possible to reduce the uncertainties in the Bell-Steinberger
relation by two orders of magnitude, which will make this CPT test be
sensitive at the Planck scale also.

     The Bell-Steinberger relation \cite{kn:bell} is a statement of the
conservation of probability in $K^0 - {\overline K^0}$ decays, in
which, through Eq. (\ref{eq:kkbar}), $\Delta$ appears.  It is usually
written as:
\begin{eqnarray}
(1+i\tan{\phi_{SW}})[Re(\epsilon)-iIm(\Delta)] = \sum_{f} \alpha_f
\label{eq:bellsteinberger}
\end{eqnarray}
where the sum runs over all decay channels f, and
$\alpha_f=\frac{1}{\Gamma_S} A^*({\rm K_S}\rightarrow f) A({\rm
K_L}\rightarrow f)$.  The most
recent published evaluation of the Bell-Steinberger relation is ref.
\cite{kn:thomson}.

     The biggest uncertainties in the Bell-Steinberger relation at
this time come from $\eta_{000}$, $Im(x)$, and $\delta_l$ (the charge
asymmetry in $K_L$ semileptonic decays).  Although $\delta_l$ doesn't
explicitly appear in the Bell-Steinberger relation, it is the best way of
evaluating $Re(\epsilon)$.  The proposed experiment will be able to
make excellent measurements of the first two of these quantities, and
KTeV will measure $\delta_l$ quite accurately.  For the next level of
accuracy in the Bell-Steinberger relation the uncertainties of the
$\alpha_{+-}$ and $\alpha_{00}$ terms must be reduced.  These
uncertainties depend on those of $|\eta_{+-}|,
Re(\epsilon'/\epsilon)$, and $\Delta \phi = \phi_{00} - \phi_{+-}$.
The latter two quantities will be measured by the KTeV experiment to
sufficient accuracy for our purposes here.

     We will have good sensitivity for the $|\eta_{+-}|$ measurement.
In our fits to the proper time dependence of $\pi^+ \pi^-$ events we
have excellent statistical sensitivity for measuring $|\eta_{+-}|$. In
the interference term, however, it is highly correlated with $D$, the
dilution factor.  We will measure $D$ using semileptonic decays. The
semileptonic charge asymmetry at zero proper time equals $D$.  We
calculate that we will be able to measure $D$ to better than 0.1\% for
momenta above 13 GeV/c.  We should be able to measure $|\eta_{+-}|$ to
0.1\% accuracy, about 10 times better than it is currently known.

     The most accurate way to determine $|\eta_{00}|$ will be by using
the KTeV value of $\epsilon'/\epsilon$ and our measurement of
$|\eta_{+-}|$.  The most accurate way of determining $\phi_{00}$ will
use the KTeV value of $\Delta \phi$ and our measurement of $\phi_{+-}$.

     We should be able to reduce the uncertainties in the
Bell-Steinberger relation by about two orders of magnitude from their
present values.  The limit on $Re(\Delta)$ will be about $5 \times
10^{-6}$, about twice the contribution of CPT violation at the Planck
scale, and will be set by the uncertainty in $\delta_l$.  For
$Im(\Delta)$ the limit will be about $1 \times 10^{-6}$, dominated by
the uncertainty in $Im(x)$, which would allow us to place a $2\sigma$
limit at the Planck scale.  Since the Bell-Steinberger measurement is
sensitive to $Re(\Delta)$ and $Im(\Delta)$ independently, these limits
would be valid even if $\Delta$ is parallel to $\epsilon$, in contrast
to the CPT violation limits from $|\phi_{+-} - \phi_{\epsilon}|$,
which are sensitive only to the component of $\Delta$ perpendicular
to $\epsilon$.

\section{Conclusion}

     We have described an experiment to carry out a systematic program
of measurements in $K_S - K_L$ interference physics.  We will search
for CPT symmetry violation in the decays of $K^0$ mesons with the
sensitivity to reach the Planck scale, measure CP violation parameters
to test the detailed predictions of the Standard Model, and study rare
kaon decays.

     Our design uses protons from the Fermilab Main Injector to make
an RF separated $K^+$ beam.  With this we make a tertiary neutral kaon
beam created in just the way to maximize the interference between
$K_S$ and $K_L$ while maintaining high flux.  We use a ``closed
geometry'' hyperon magnet for beam definition.  A standard Vee
spectrometer, with drift chambers, an electromagnetic calorimeter, and
a muon detector, is used to make the measurement.


\begin{thebibliography}{99}
\bibitem{kn:stcomment}  The CPT theorem is invalid in string theory
because strings are nonlocal objects.
\end{thebibliography}
\end{document}